# A Comprehensive Study of Commonly Practiced Heavy & Light Weight Software Methodologies

[1]Asif Irshad Khan, [2]Rizwan Jameel Qurashi and [3]Usman Ali Khan

[1] Department of Computer Science
Singhania University, Jhunjhunu, Rajasthan, India

2 Department of Information Technology
King Abdul Aziz University, Jeddah, Saudi Arabia

[3] Department of Information System
King Abdul Aziz University, Jeddah, Saudi Arabia

## Abstract

Software has been playing a key role in the development of modern society. Software industry has an option to choose suitable methodology/process model for its current needs to provide solutions to give problems. Though some companies have their own customized methodology for developing their software but majority agrees that software methodologies fall under two categories that are heavyweight and lightweight. H eavyweight methodologies (Waterfall Model, Spiral Model) are also known as the traditional methodologies, and their focuses are detailed documentation, inclusive planning, and extroverted design. Lightweight methodologies (XP, SCRUM) are, referred as agile methodologies. Light weight methodologies focused mainly on short iterative cycles, and rely on the knowledge within a team.

The aim of this paper is to describe the characteristics of popular heavyweight and lightweight methodologies that are widely practiced in software industries. We have discussed the strengths and weakness of the selected models. Further we have discussed the strengths and weakness between the two opponent methodologies and some criteria is also illustrated that help project managers for the selection of suitable model for their projects.

## 1. Introduction

Software development life cycle (SDLC) describes the life of a software product in terms of processes from its conception to its development, implementation, delivery, use and maintenance. Software development usually involves following stages (processes) as shown in figure 1. A software process is a set of activities that lead to the production of software product. Processes are also called as methodologies helping to maintain a level of consistency and quality on a set of activities. A process is a collection of procedures that are used to a p roduct by meeting a set of goals or standards.

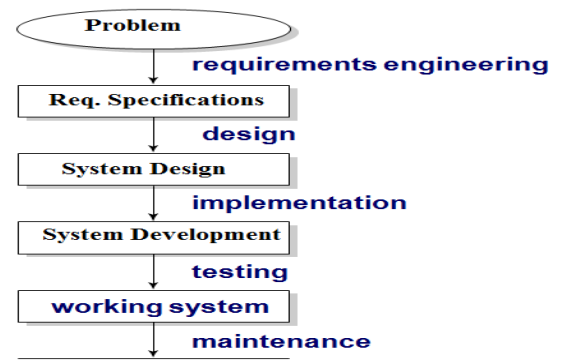

Fig.1 Stages of SDLC

Many process models are described in the literature such as Waterfall, Prototype, Rapid Application development (RAD), Spiral Model, Object Oriented, Agile and Component Based Development. Following are software improvement factors.

- Development speed (time to market)
- Product quality and Project visibility
- Administrative overhead
- Risk exposure and customer relations, etc.

The paper is organized as: **section 2** describes some selected Heavy weight Models along their advantages and disadvantages, **section 3** discussed some selected Light weight models along their strengths and weaknesses, **section 4** Comparison of Heavy and Light weight models in terms of differences and issues and **section 5** discussed criteria for the selection of process models to develop a project.

## 2. Heavy Weight Methodologies

Heavyweight development methodology is based on a sequential series of steps, such as requirements definition, solution build, testing and deployment. Heavyweight development methodology mainly focuses detailed documentation, inclusive planning, and extroverted design. Following are the most popular Heavyweight development methodologies.

### 2.1 Spiral Model

This model is also known as Boehm's model, using his model, process is represented as a spiral rather than as a sequence of activities with backtracking. The spiral has many cycles and each cycle represents a phase in the process. No fixed phases such as specification or design – Loops, In the Spiral model are chosen depending on what is required. Risks are explicitly assessed and resolved throughout the process. The structure of the Spiral model is shown in the figure -2.

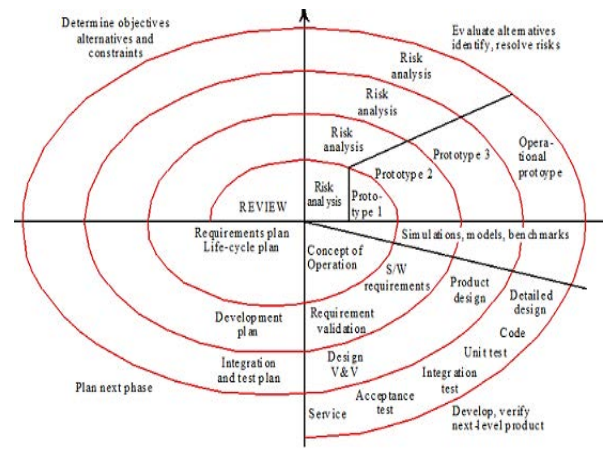

Fig.2 Spiral model [2]

Following are the four main phases of the spiral model:

**Objective setting** – Specific objectives for the project phase are identified.
**Risk assessment and reduction** – Key risks are identified, analyzed and information is obtained to reduce these risks. The aim is that all risks are resolved.
**Development and Validation** – Once all possible risks have been identified the development of the software can begin, an appropriate model is chosen for the next phase of development.
**Planning** – The project is reviewed and plans are drawn up for the next round of spiral.

### Usage of Spiral Model
Following are the usage of Spiral model.
- Large and high budget projects.
- When risk assessment is very critical.
- Requirements are not very clearly defined and complex.
- When costs and risk evaluation is important
- For medium to high-risk projects.
- Long-term project commitment unwise because of potential changes to economic priorities.

### 2.2 Rational Unified Process Model (RUP)
The Rational Unified Process (RUP), originated by Rational Software and later by IBM, is a heavy iterative approach that takes into account the need to accommodate change and adaptability during the development process. In RUP: a software product is designed and built in a succession of incremental iterations. Each iteration includes some, or most, of the development disciplines like requirements, analysis, design, and implementation and testing. It provides a disciplined approach to assign tasks and responsibilities within a software development organization for the successful development of software [1,2].

The main goal of RUP is to ensure the production of high-quality software by meeting the needs of its end-users within a predictable schedule and budget. The Rational Unified Model provides guidelines, templates and tool mentors to software engineering team to take full advantage of among others. Following are the guidelines for best practices:

1. Develop software iteratively
2. Manage requirements
3. Use component-based architectures
4. Visually model software
5. Verify software quality
6. Control changes to software

RUP has a series of phases as follows.

**Inception:** (**Understand what to build**) In this phase project's scope, estimated costs, risks, business case, environment and architecture are identified.

**Elaboration: (Understand how to build it)** In this phase requirements are specified in detail, architecture is validated, the project environment is further defined and the project team is configured.

**Construction: (Build the Product)** In this phase software is built and tested and supporting documentation is produced.

**Transition: (Transition the Product to its Users)** In this phase software is system tested, user tested, reworked and deployed.

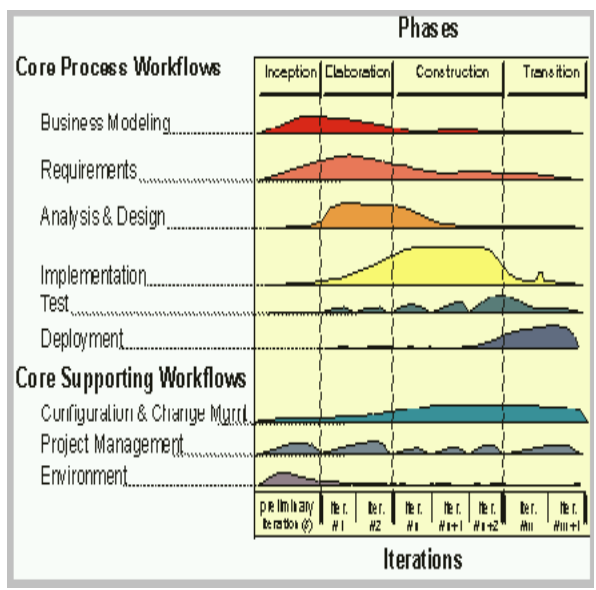

Fig.3 Phases of RUP model [3]

**Usage of RUP model**
Following are the usage of the RUP model
a) Distributed systems
b) Very large or complex systems
c) Systems combining several business areas
d) Systems reusing other systems
e) Distributed development of a system

## 2.3 Incremental Model

Incremental model is an evolution of Waterfall model. The product is designed, implemented, integrated and tested as a series of incremental builds. The releases are defined by beginning with one small, functional subsystem and then adding functionality with each new release. In the incremental model, there is a good chance that a requirements error will be recognized as soon as the corresponding release is incorporated into the system. With the Incremental model, portions of the total product might be available within weeks, whereas the client generally waits months or years to receive a product built using the Waterfall model. The gradual introduction of the product via the incremental model saves time of client.

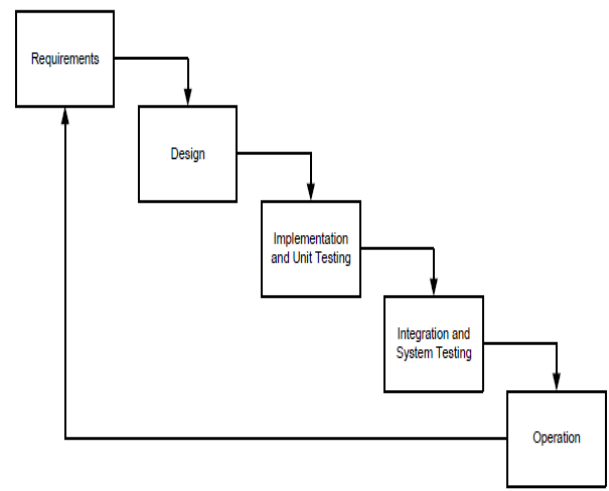

Fig.4 Incremental Model is a Series of Waterfalls [3]

**Usage of Incremental Model**
The incremental model is good for projects where requirements are known at the beginning, it is best used on low to medium-risk programs. If the risks are too high to build a successful system using a single waterfall cycle, spreading the development out over multiple cycles may lower the risks to a more manageable level.

## 2.4 Component Based Development Model

One of the main principles of computer science, divide and conquer the bigger problem into smaller chunks to solve it, fits into component based development. The aim is to build large computer systems from small pieces called a component that has already been built instead of building complete system from scratch.

A software component technology is the implementation of a component model by means of:
- Standards and guidelines for the implementation and execution of software components.
- Software tools that supports the implementation, assembly and execution of components.

When we look at the lifecycle of a component, it passes through the following global stages requirement analysis, design, development, packaging, testing, distribution, deployment and execution.

Preliminary design entails component specification, while detailed design consists of component search and identification. For each application is constructed, the developers need to manage all the components used, along with their version information. As developers frequently release multiple versions of an application, it's important that they manage component versions used in applications [4]

In the development stage, the design, specification, implementation and meta-data of components is constructed.

In the packaging stage, all information that is needed for trading and deployment of the component implementation grouped into a single package.

The distribution stage deals with searching, retrieval and transportation of components. For searching components meta-data is needed that need to be included in the packaging stage.

The deployment stage address issues related to the integration of component implementations in an executable system on some target platform. Finally, the execution stage deals with executing and possibly upgrading components.

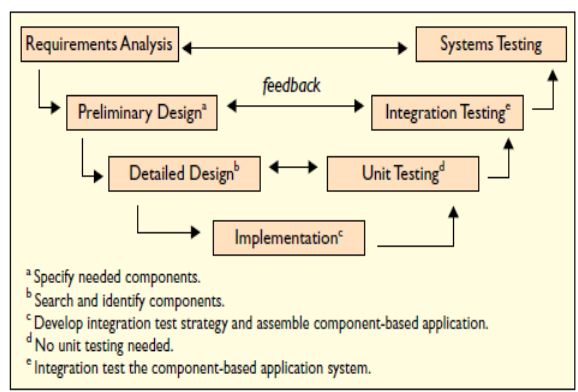

Fig.5. Component-Based Application Integration Life Cycle [4]

Table 1: Advantages and Disadvantages of some Heavy Weight Methodologies

| Advantages | | | |
|---|---|---|---|
| **Spiral Model** | **RUP Model** | **Incremental Model** | **Component Model** |
| Emphasize planning for verification and validation of the product in early stages of product development. | The iterative approach leads to higher efficiency. | With every increment operational product is delivered. Early increments can be implemented with fewer people. | Reduction of cost through reuse of previously developed assets in the development of new Systems. |
| Each deliverable must be testable. | Testing takes place in each iteration, not just at the end of the project life cycle. This way, problems are noticed earlier, and are therefore easier and cheaper to resolve. | Testing and debugging during smaller iteration is easy. | In principle, more reliable systems, due to using previously tested components. |
| Good for large and mission-critical projects. | Managing changes in software requirements will be made easier by using RUP, i.e. Change is more manageable. | More flexible – less costly to change scope and requirements. | Facilitating the maintenance and evolution of systems by Easy replacement of obsolete components with new enhanced ones. |
| Software engineers can | The development time required is | Easier to manage risk | This Model brings the |

| get their hands in and start working on a project earlier. | less due to reuse of components. | because risky pieces are identified and handled during its iteration. | potential for enhancing the quality of enterprise software systems. |
|---|---|---|---|
| **Disadvantages** | | | |
| **Spiral Model** | **RUP Model** | **Incremental Model** | **Component Model** |
| Does not easily handle concurrent events and iterations. | Not suitable for small scale industry and safety critical projects. | Each additional build has to incorporate into the existing structure without degrading the quality of what has been released earlier. | Difficulty in locating suitable component. |
| Doesn't work well for smaller projects. | This Model is too complex, too difficult to learn, and too difficult to apply correctly if you don't have an expert project managers or project members. | The planning the delivery increments are critical to the success. Wrong planning can result in a disaster. Not suitable for large, long-term projects. | Problem in understanding such component. |
| Does not easily handle dynamic changes in requirements. | On cutting edge projects which utilize new technology, the reuse of components will not be possible. | Clients are required to learn how to use a new system with each deployment. | Conversion cost to a "reuse situation" (tools, training, culture) |
| Highly customized limiting re-usability. | RUP is a commercial product, no open or free standard. Before RUP can be used, the RUP has to be bought from IBM, which can sometimes limit its use. | Design issues may arise because not all requirements are gathered. | |

## 3. Light Weight Methodologies

Light weight development methodologies embrace practices that allow programmers to build solutions more quickly and efficiently, with better responsiveness to changes in business requirements. Light weight methodology mainly focuses development, based on short life cycles, involves the customer, Strive for simplicity and value the people. Following are some of the popular lightweight development methodologies.

### 3.1 Agile Process Model

With the rapid change in the requirements in terms of budget, schedule, resources, team and technology agile model responds to changes quickly and efficiently. Agile is an answer to the eager business community asking for lighter weight along with faster and nimbler software development processes [7].

Following are the main principles to implement an agile model.

(1) Agile team and customer must communicate through face-to-face interaction rather than through documentation.

(2) Agile team and customer must work together throughout the development.

(3) Supply developers with the resources they need and then trust them to do their jobs well.

(4) Agile team must concentrates on responding to change rather than on creating a plan and then following it.

(5) Emphasis on good design to improve quality.

(6) Agile team must prefer to invest time in producing working software rather than in producing comprehensive documentation.

(7) Satisfy the customer by "early and continuous delivery of valuable software".

**Extreme Programming (XP)**

Extreme Programming (XP) is a lightweight design method developed by Kent Beck, Ward Cunningham, and others, XP has evolved from the problems caused by the long development cycles of traditional development models. The XP process can be characterized by short development cycles, incremental planning, continuous feedback, reliance on communication, and evolutionary design. With all the above qualities, XP programmers respond to changing environment with much more courage [5,6].

A summary of XP terms and practices is listed below:
**Planning** – The programmer estimates the effort needed for implementation of customer stories and the customer decides the scope and timing of releases based on estimates.

**Small/short releases** – An application is developed in a series of small, frequently updated versions. New versions are released anywhere from daily to monthly.
**Metaphor** – The system is defined by a set of metaphors between the customer and the programmers which describes how the system works.
**Simple Design** – The emphasis is on designing the simplest possible solution that is implemented and unnecessary complexity and extra code are removed immediately.
**Refactoring** – It involves restructuring the system by removing duplication, improving communication, amplifying and adding flexibility but without changing the functionality of the program.

a) **Pair programming** – All production code are written by two programmers on one computer.
b) **Collective ownership** – No single person owns or is responsible for individual code segments rather anyone can change any part of the code at any time.
c) **Continuous Integration** – A new piece of code is integrated with the current system as soon as it is ready. When integrating, the system is built again and all tests must pass for the changes to be accepted.
d) **40-hour week** – No one can work two overtime weeks in a row. A maximum of 40-hour working week otherwise it is treated as a problem.
e) **On-site customer** – Customer must be available at all times with the development team.
f) **Coding Standards** – Coding rules exist and are followed by the programmers so as to bring consistence and improve communication among the development team.

**Usage of XP model** Best for projects that are not large, and involve a small group of developers working for a limited time period. But agile processes require that the developers involved be highly skilled.

**Scrum** The Scrum Methodology is based on the Rugby term for individual groups collaborating together to form a powerful whole. In Scrum, projects are divided into brief work cadences, known as sprints. Each sprint is typically one week to four weeks in duration. The Sprints are of fixed duration and never extended, each of which results in a potentially usable product with an added increment of function.

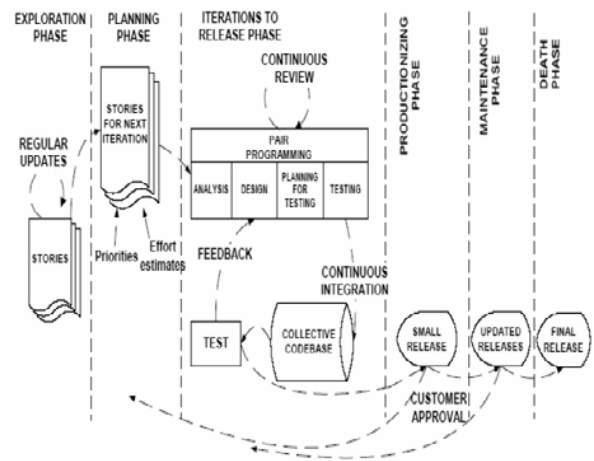

Fig.6 Lifecycle of the XP Process [5]

The tasks for each sprint are set, in consultation with a stakeholder representative, during a sprint planning meeting and cannot be added to during the sprint. Each task is typically expressed as a user story. Tasks that are not able to accomplish in time are returned by the team to the backlog for future consideration. The structure of the scrum model is shown in the figure 7

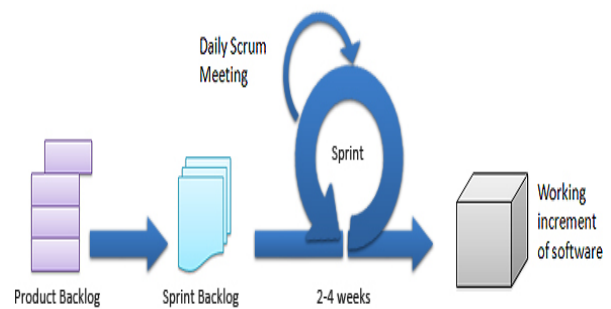

Fig.7 Scrum Software Development Model [7]

### 3.2 Prototype Model

Is a p artially developed product that enables customers and developers to examine some aspect of the proposed system and decide if it is suitable or appropriate for the finished product.

For example developers may build a p rototype system for key requirements to ensure that the requirements are feasible and practical and discussed with the customers by showing the prototype system, if not, revisions are made at the requirements stage rather than at the more costly testing stage.

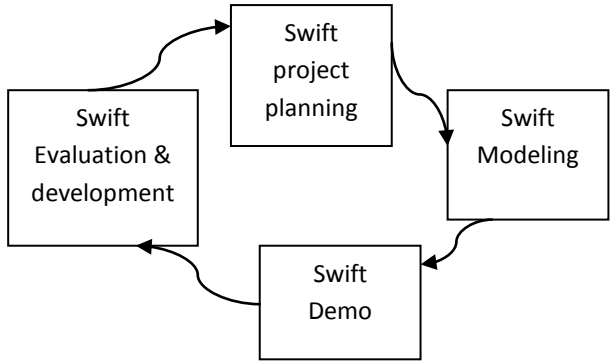

Fig.8 Prototype Model [2]

### 3.2 Rapid Application Development (RAD) Model

Rapid Application Development (RAD) is an incremental software development process model that emphasises a very short development cycle and encourages constant feedback from customers throughout the software development life-cycle. The main objective of Rapid Application Development is to avoid extensive pre-planning, generally allowing software to be written much faster and making it easier to change requirements.

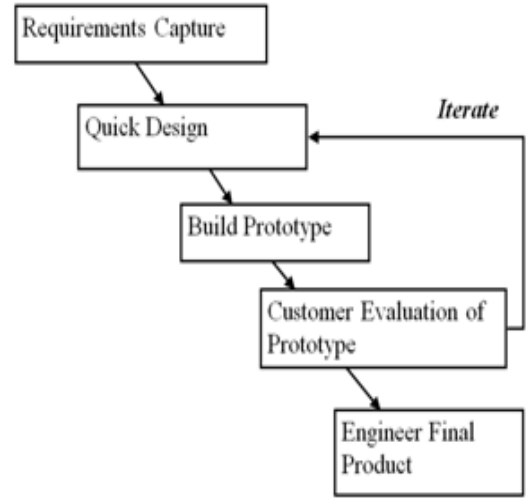

Fig.9 RAD prototype model [2]

Strength and weakness of some of the popular light weight methodologies are listed in Table 2.

**Table 2- Strength and weakness of some Light Weight Methodologies**

| Strength of some Light Weight Methodologies | | | |
|---|---|---|---|
| **Prototype Model** | **(RAD) Model** | **(XP) Model** | **Scrum Model** |
| Faster development results in early delivery and cost saving. | Time to deliver is less. | XP model suit small-medium size projects. It mainly emphasis on customer involvement: A major help to projects where it can be applied. | It provides an open forum, where everyone knows who is responsible for which item. |
| Can integrate with other models such as waterfall to produce effective results. | Quick development results in saving of time as well as cost. | Emphasis on good team cohesion. | Focus on team communications, Team spirit and solidarity. |
| Improved and increased user involvement. | Productivity with fewer people in short time. | Emphasizes final product. | Frequent demonstrations for early feedback from stakeholders. |
| . | Progress can be measured. | Test based approach to requirements and quality assurance. | Scrum is mainly useful for fast moving web 2.0 or new media projects. |

| Weakness of some Light Weight Methodologies ||||
|---|---|---|---|
| **Prototype Model** | **(RAD) Model** | **(XP) Model** | **Scrum Model** |
| Lack of documentation results costly support for upgrading of the software | Suitable only when requirements are well known. | Difficult to balance up to large projects where documentation is essential. | Decision-making is entirely in the hands of the teams. |
| Rapid development often results in poor quality software | Requires user involvement throughout the life cycle. | Needs experience and skill if not to degenerate into code-and-fix. | To deliver the project in time there is always a need of experienced team member's only. |
| Concentrate mainly on experimenting with the customer requirement may results in poorly understood. | Only Suitable for project requiring shorter development times. | Lack of design documentation also programming pairs is costly. | Project development may be effected hugely, If any of the team members leave during development. |
| It is a big cost saver in terms of project budget as well as project time and cost due to reusability of the prototypes. | Product may lose its competitive edge because of insufficient core functionality and may exhibit poor overall quality. | The XP method provides essentially no data-gathering guidance.  XP does not explicitly plan, measure, or manage program quality. | Present of Uncommitted team members may results in project failure. |
|  | Success depends on the extremely high technical skilled developers. | Methods are only briefly described, when the XP method fails in practice, this is usually the cause. | Scrum requires a certain level of training for all users, this can increase the overall cost of the project. |

## 4: Comparison of Heavy and Light Weight Models in terms of differences and issues

Comparison of Heavy and Light Weight Models in terms of differences and issues are listed in Table 3 [5,8].

### Table 3- Comparison of Heavy and Light Weight Models

| **Differences** | **Light weight** | **Heavy weight** |
|---|---|---|
| Customers | Committed, knowledgeable, collocated, collaborative, representative, empowered | Access to knowledgeable, collaborative, representative, empowered customers |
| Requirements | Mainly emergent, rapid change. | Knowable early, largely stable. |
| Architecture | Designed for current requirements | Designed for current and foreseeable requirements |
| Size | Smaller Team and Products | Larger Team and Products |
| Primary objective | Rapid value | High assurance |
| Developers | Knowledgeable, collocated, collaborative | Plan-driven, adequate skills, access to external knowledge. |
| Release cycle | In phases (multiple cycles) | Big bang (all functionality at once) |
| **Issue** | **Light Weight** | **Heavy Weight** |
| Development life cycle | Linear or incremental | incremental |
| Style of development | Adaptive | Anticipatory |
| Requirements | Emergent – Discovered during the project. | Clearly defined and documented |

| Documentation | Light (replaced by face to face communication) | Heavy / detailed Explicit knowledge |
|---|---|---|
| Team members | Co-location of generalist senior technical staff | Distributed teams of specialists |
| Client Involvement | onsite and considered as a team member Active/proactive | Low involvement |
| Market | Dynamic/Early market | Mature/Main Street market |
| Measure of success | Business value delivered | Conformance to plan |

## 5. Criteria for the Selection of Process Model

Selection of the suitable software model to use within an organization is critical for overall success of the project. A project Manager can use the following criteria to select a suitable model for the development of a new project as per needs.

The selection of one model over the others is driven by Project size, Budget, Team size, criticality of the project and a lot of other factors. The different aspects which need to be kept in mind while selecting a suitable process model can be summarized as follows:-

- Reuse of existing Components? Component Based approach may be ideal choice.
- Very large project with High risks or high cost of failure? A Spiral process Model may be your best choice.
- Although your customer have defined business goals but the requirement are not freeze yet Agile (light Weight) Model will have the advantage over others as it has the flexibility to change the requirement at any stage.
- All the developers are experts? And if the project is small enough, an agile approach may work for you.
- Want to keep stakeholders involved? An Incremental process Model may be what you need.
- Don't have an on-site stakeholder to sit with the developers? Agile development model is not for you.
- Developers are not highly skilled? A Waterfall or Spiral process Model can help keep them on track and out of trouble.

## 6. Conclusion

No one model is necessarily better or worse than another. As always the selection of a particular model is correct only in the context of the organization or the product under development. Correct selection is to a great degree dependant on having a clear understanding of the groups and types of development models. This is of further importance given that it's highly unlikely that any organization will follow a model strictly. Most organizations will opt to use a hybrid form that fits the capabilities of their staff and meets the needs of their business.

A. A. Name, and B. B. Name, Book Title, Place: Press, Year.
[2] A. Name, and B. Name, "Journal Paper Title", Journal Name, Vol. X, No. X, Year, pp. xxx-xxx.
[3] A. Name, "Dissertation Title", M.S.(or Ph.D.) thesis, Department, University, City, Country, Year.
[4] A. A. Name, "Conference Paper Title", in Conference Name, Year, Vol. x, pp. xxx-xxx.

## References

[1] M. Fowler, "The New Methodology," Available at http://www.martinfowler.com/articles/newMethodology.html Accessed on 25/03/2011.
[2] Ian Sommerville, Software Engineering, UK: Addison Wesley, 2004.
[3] Julien Lemétayer "identifying the critical factors in software development methodology FIT", Victoria University of Wellington.
[4] Padmal Vitharana, "risks and challenges of component-based software development" communications of the ACM, Vol. 46, No. 8, August 2003, pp. 67-72.
[5] NK. Beck, Extreme Programming explained: Embrace change. Reading, Mass., USA: Addison-Wesley, 2004.
[6] M. Rizwan Jameel Qureshi and S.A Hussain, "An Adaptive Software Development Process Model", Advances in Engineering Software, Elsevier Ltd Amsterdam, The Netherlands, Vol.39, No. 8, 2008, pp. 654-658.
[7] "Scrum Agile Model" available at: http://www.rightwaysolution.com/scrum-agile-development-model.html accessed on: 24/04/2011
[8] M. Rizwan Jameel Qureshi and S.A Hussain, "A Reusable Software Component-Based

Development Process Model", Advances in Engineering Software, Elsevier Ltd, Amsterdam, The Netherlands**.** Vol. 39, No. 2, 2008, pp. 88-94.

**Asif Irshad Khan** received his bachelor and Master degree in Computer Science from the Aligarh Muslim University (A.M.U), Aligarh, India in 1998 and 2001 respectively. He is presently working as a Lecturer Computer Science at Faculty of Computing and Information Technology, King Abdul Aziz University, Jeddah, Saudi Arabia. His area of interest include software engineering, component based software engineering and object oriented technologies. He is a **Ph. D. scholar in Singhania University, Pacheri Bari, Dist. Jhunjhunu, Rajasthan, India enrolled in 2010.**

**Dr. M. Rizwan Jameel Qureshi**, is an assistant Professor at IT Department, Faculty of Computing and Information Technology, King Abdul Aziz University, Jeddah, Saudi Arabia. He has done his Ph.D. CS (Software Process Improvement) in 2009. He is in the field of teaching and research since 2001. He has published sixteen research papers and five books at national and international forums. He is teaching Software Engineering domain courses at graduate and undergraduate level from more than ten years.

**Dr. Usman Ali Khan**, is an assistant Professor at IS Department, King Abdul Aziz University, Jeddah, Saudi Arabia. He is in the field of teaching and research since 1995. He has published thirteen research papers at national and international forums. He is teaching Software Engineering domain courses at graduate and undergraduate level from more than ten years.